# Improved transmission method for measuring the optical extinction coefficient of micro/nano particle suspensions

X. C. Li,[1] J. M. Zhao,[1,*] C. C. Wang,[1,2] L. H. Liu [1,3,*]

[1] *School of Energy Science and Engineering, Harbin Institute of Technology, Harbin 150001, China*
[2] *School of Automobile Engineering, Harbin Institute of Technology at Weihai, Weihai 264209, China*
[3] *Department of Physics, Harbin Institute of Technology, Harbin 150001, China*
*Corresponding author: jmzhao@hit.edu.cn (J.M. Zhao), lhliu@hit.edu.cn(L.H. Liu)*

**Extinction coefficient is fundamental to analyze radiative transport in micro/nano particle suspensions. In the traditional transmission method for measuring the extinction coefficient of particles in a cuvette, a reference system is used to compensate the influence of the cuvette and base fluid. However, the multiple reflections and refractions between the air/glass and liquid/glass interfaces cannot be sufficiently eliminated by using the reference system, and the induced measurement error increases significantly with increasing difference in refractive index between two neighboring media at these interfaces. In this paper, an improved transmission method is proposed to measure the extinction coefficient of micro/nano particles. The extinction coefficient of the particles is determined based on an optical model taking into account the multiple reflection and refraction at the glass/liquid interfaces. An experimental validation was conducted for suspensions with various mean particle sizes. By considering the higher-order transmission terms, the improved transmission method generally achieved high accuracy improvement over the traditional transmission method for extinction coefficient measurement, especially for the case with small optical thickness of particle suspensions. This work provides an alternative and more accurate way for measuring the extinction characteristics of micro/nano particle suspensions.**

http://dx.doi.org/10.1364/AO.99.099

## 1. INTRODUCTION

The micro-/nano- sized particles are widely used in numerous fields, including material science, chemical engineering, biotechnology, and atmospheric aerosol science, etc [1-3]. In recent years, the micro-/nano-particles have attracted significant interests for their potential applications in solar energy harvesting [4-11]. The knowledge of the radiative properties, such as extinction, absorption and scattering coefficients, and the scattering phase function, is crucial to analyze radiative transfer in the liquid-particle suspensions [12-14].

The optical extinction characteristics of liquid suspensions of micro-/nano- sized particles have been investigated by many researchers. When the diameters of particles are far less than the wavelength λ, say, λ/10 or less, the effective medium approach (EMA) can be used and the radiative properties of the suspensions were characterized by the effective optical constants [15, 16]. Taylor et al. [17] proposed a method to determine the extinction coefficient of nanofluids based on the EMA. Kameya et al. [4] measured the radiation absorption characteristics of a Ni nanoparticle suspension by spectroscopic transmission measurement for two pathlengths using optical models based on the EMA. When the particle sizes are larger, however, the particle scattering cannot be neglected and the EMA will not work.

Generally, the extinction characteristics of particles are determined from normal–normal transmittance measurements based on Beer-Lambert's law, and a reference system is used to compensate the influence of the cuvette and base fluid. Figure 1 shows the schematic diagram of the light transmission. Here, $q_0$ is the incident collimated light, and the transmitted light flux contains the first-order transmission, second-order transmission and so on. However, the simplified optical model used in the traditional approach to obtain the extinction coefficient omits higher order transmission. The use of a reference system is not sufficient to eliminate the effect of higher order transmissions (even the second order cannot be eliminated). The measured transmittances of the glass–liquid suspensions of particles–glass system $T_{\text{EXP}}$ and the glass–base fluid–glass system (reference system) $T_{\text{EXP,Ref}}$ are expressed respectively as

$$T_{EXP} = \underbrace{\left[T_1 e^{-\beta L_2} T_3\right]}_{\text{1st order}} + \underbrace{\left[R_1' T_1 \left(e^{-\beta L_2}\right)^3 T_3 R_3\right]}_{\text{2nd order}} + ... \quad (1)$$

$$T_{EXP,Ref} = \underbrace{\left[T_1 e^{-\beta_{base} L_2} T_3\right]}_{\text{1st order}} + \underbrace{\left[R_1' T_1 \left(e^{-\beta_{base} L_2}\right)^3 T_3 R_3\right]}_{\text{2nd order}} + ... \quad (2)$$

where $e^{-\beta L_2}$ and $e^{-\beta_{base} L_2}$ are the transmissivities of the liquid suspensions of particles and the base fluid, respectively. $T_1$ and $T_3$ represent the transmittance of the left wall and right wall of cuvette, respectively. $\beta$ and $\beta_{base}$ are the extinction coefficients of the liquid-particle suspensions and the base fluid, respectively. $L_2$ is the thickness of the liquid layer. Note that a close form of the summation series Eq. (1) can also be obtained as given by Eq. (11).

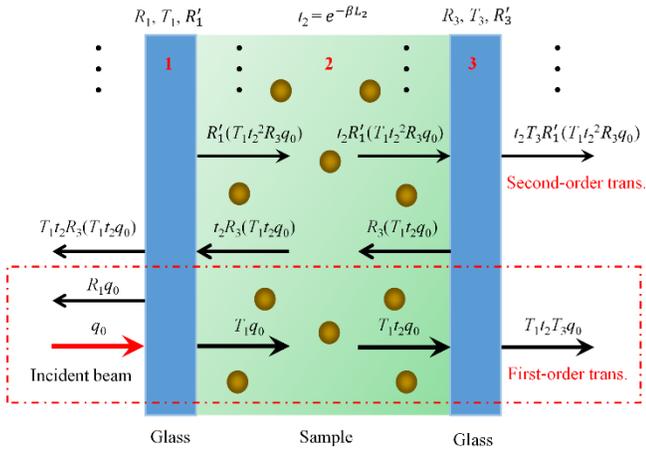

**Fig. 1**. Schematic diagram of the light transmission. $T_1$ and $T_3$ represent the transmittance of layer 1 and 3 (glass), respectively. $t_2$ represents the transmissivity of layer 2 (liquid-particle suspensions), $R_1$ and $R_1'$ represent the reflectance of layer 1 from the incident side and from the non-incident side, respectively. $R_3$ and $R_3'$ represent the reflectance of layer 3 from the incident side and from the non-incident side, respectively.

In the traditional transmission method, the apparent extinction coefficient of particles $\beta_{particle}^{trad}$ is written as [2, 18-20]

$$\beta_{particle}^{trad} = -\frac{1}{L_2} \ln\left(\frac{T_{EXP}}{T_{EXP,Ref}}\right) \quad (3)$$

It is noted that Eq. (3) cannot be obtained if the nonlinear higher order transmission terms are not omitted. Hence Eq. (3) is an approximate relation. By using Eqs. (1) and (2), the relation between $\beta_{particle}^{trad}$ with the true value of extinction coefficient $\beta_{particle}$ can be derived as

$$\beta_{particle}^{trad} = \beta_{particle} + \frac{1}{L_2} \ln\left(\frac{1 - R_1' R_3 t_2^2}{1 - R_1' R_3 t_{2,base}^2}\right) \quad (4)$$

where $t_2 = e^{-\beta L_2}$, $t_{2,base} = e^{-\beta_{base} L_2}$. Note that the last term in Eq. (4) is the theoretical truncation error of extinction coefficient determined using the traditional method (Eq. (3)).

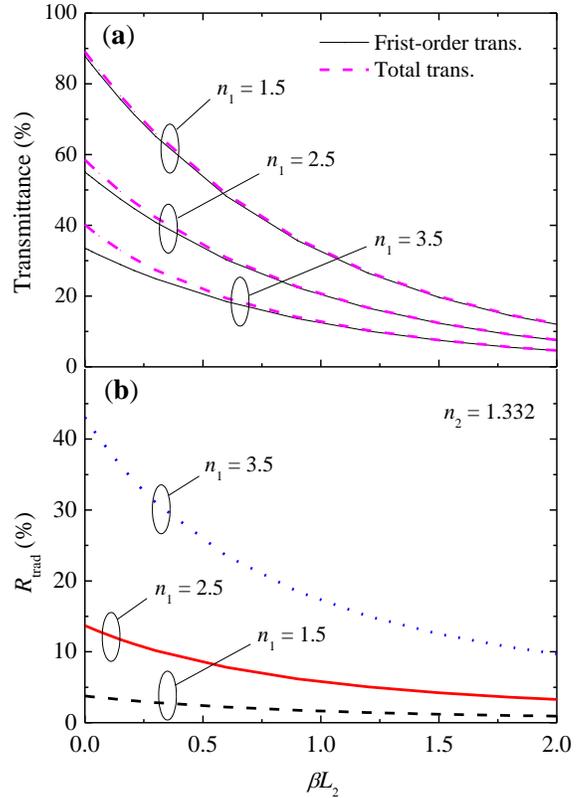

**Fig. 2.** (a) Comparison of transmittance obtained using the first-order transmission and the total transmission and (b) theoretical relative error of the extinction coefficient determined using the traditional approach ($R_{trad} = \left|\beta_{particle}^{trad} - \beta_{particle}\right|/\beta_{particle}$) as a function of optical thickness of sample liquid at different refractive indices of glass.

As shown in Fig. 1, the optical model used in the traditional transmission method does not consider higher-order transmission terms. For further understanding of the effect of higher-order transmission terms on the accuracy of extinction coefficient measurement, a numerical analysis of the first-order and the total transmission terms is conducted. In the analysis, the optical thickness $\beta L_2$ of the sample medium was varied from 0 to 2.0, the measured transmittance was simulated using the three-layer system (glass–sample–glass) model based on Fig. 1. The thickness of glass is set as 1.5 mm. The refractive index of sample medium is $n_2 = 1.332$. Figure 2 shows the variation of transmittance obtained using the first-order transmission and the total transmission and theoretical relative error of the extinction coefficient determined using the traditional approach as a function of optical thickness of sample liquid at different refractive indices of glass. The difference between the first-order transmission and total transmission becomes larger when the optical thickness is smaller. For example, in the case of $\beta L_2 = 0.15$, the differences of transmittance are about 1%, 5% and 14% for refractive indices of glass $n_1 = 1.5, 2.5,$ and 3.5, respectively. The selection of higher value of refractive index of glasses refers to window materials such as diamond, ZnSe, ZnS, and Si, etc., which are usually used in high temperature, high pressure or infrared measurement. Theoretical relative error of the extinction coefficient determined using the traditional approach $R_{trad}$ increases with the decreasing of optical thickness. When $\beta L_2 = 0.15$, $R_{trad}$ are about 3.5%, 12% and 36% for $n_1$ = 1.5, 2.5, and 3.5, respectively. It is observed that the errors will be more significant if the cuvette is made of higher refractive index glass, which is due to higher reflection at interfaces. The theoretical relative error of extinction coefficient determined using traditional method is about 3 times of the related relative error of transmittance when higher order transmission is omitted, indicating the accuracy of determined extinction coefficient is very sensitive to the error of transmittance measurement for the traditional method. Furthermore, as will be experimentally demonstrated in Section 4, the uncertainty of extinction coefficient will be significantly enlarged further due to the uncertainties of transmittance measurement, both in $T_{EXP}$ and in $T_{EXP,Ref}$.

In the present work, an improved approach is proposed to measure the extinction coefficient of micro/nano particles, in which an optical model that takes into account all the higher-order transmission terms is employed. Theoretical and experimental comparison of the accuracy of the improved transmission method with that of the traditional transmission method is presented. Silicon dioxide microspheres with known optical constants and particle diameter distribution are taken as examples to verify this new method for measuring the extinction characteristics of particles.

## 2. INFLUENCE OF FORWARD SCATTERING PHOTONS

Light attenuation in liquid suspensions of particles is from the combined contributions of particles and base fluid. When the volume fraction of particle is very small, the apparent total extinction coefficient $\beta$ can be written as [17, 21, 22]

$$\beta = \beta_{particle} + \beta_{base} - \varepsilon\sigma_{particle} \tag{5}$$

where $\beta_{particle}$ and $\beta_{base}$ denote the extinction coefficient of the particles and base fluid, respectively, $\alpha_{particle}$ denotes the absorption coefficient of the particles, $\sigma_{particle}$ denotes the scattering coefficient of the particles, $\varepsilon$ is a forward-scattering-peak correction coefficient accounting for the finite size of detector acceptance angle, defined as

$$\varepsilon = \frac{1}{2}\int_0^{\theta_c}\Phi(\theta)\sin\theta\, d\theta \tag{6}$$

where $\Phi(\theta)$ is the scattering phase function of particles, $\theta_c$ denotes the half acceptance angle of the detector. The influence of forward and multiple light scattering on measurement of beam attenuation were analyzed and discussed in the published paper [23-26], indicating the influence of forward scattering is significant and cannot be omitted in case the acceptance angle of detector is large.

In reality, the scattering phase function of particles is difficult to be accurately measured, especially for a wide range of spectra, it is thus expected that the term related to forward-scattering-peak correction can be neglected with specific experiment configuration during the measurement of extinction coefficient. Then a very simple relation between the apparent total extinction coefficient and the particle extinction coefficient can be obtained, namely, $\beta = \beta_{particle} + \beta_{base}$. In the following, an analysis of the influence of forward scattering photons on the measurement accuracy of extinction coefficient is presented. A collimated beam with a light spot area $\Delta A$ irradiating normally on a layer of absorption and scattering media is considered, as shown in Fig. 3. Light transfer through the layer is governed by the steady-state radiation transfer equation (RTE) expressed in terms of the light radiation intensity $I(x,\mu)$ at location $x$, in direction $\mu$ as

$$\mu\frac{dI(x,\mu)}{dx} + \beta I(x,\mu) = \frac{\sigma}{2}\int I(x,\mu')\Phi(\mu,\mu')d\mu' \tag{7}$$

where $\beta$ and $\sigma$ are the extinction coefficient and scattering coefficient of the liquid-particle suspensions, and $\mu = \cos\theta$.

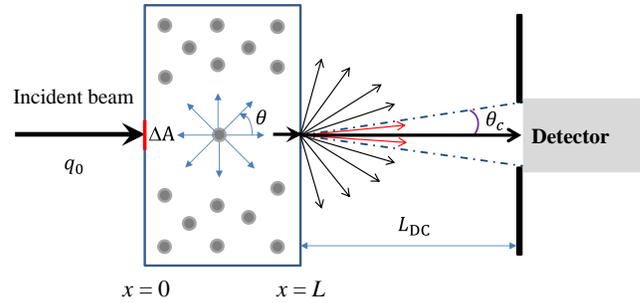

**Fig. 3.** The schematic of light forward scattering for different half acceptance angle (different distance between detector and cuvette).

According to Ref. [27], for the problem about collimated radiation impinging into the absorption and scattering media, we can separate the radiation energy within medium into two parts: (a) the remnant of the collimated beam after parted extinction along its path

$$q_c(x) = q_0 e^{-\beta x} \tag{8}$$

and (b) a diffuse part, which obeys the following equation of transfer

$$\mu\frac{dI(x,\mu)}{dx} + \beta I(x,\mu) = \frac{\sigma}{2}\int I_d(x,\mu')\Phi(\mu,\mu')d\mu' + \frac{\sigma}{4\pi\Delta A}q_0 e^{-\beta x}\Phi(\mu,1) \tag{9}$$

where $q_0$ denotes the incident light flux. Assuming the detector surface is normal with the incident light beam, the detected light flux by detector can be calculated from

$$q_{detect} = q_0 e^{-\beta L} + 4\pi\Delta A\int_0^{\cos\theta_c}I_d(L,\mu')\mu'\,d\mu' \tag{10}$$

To improve the accuracy of extinction coefficient measurement, we should eliminate the effect of diffuse radiation (the end term in Eq. (10)). From Fig.3, for a detector with fixed active area, the detecting solid angle is decreased with increasing the distance between cuvette and the detector. As seen, the contribution from the diffuse intensity is proportional to the acceptance solid angle of the detector, which can be ignored when the acceptance solid angle is chosen small enough. In this condition, the detected light flux $q_{detect}$ is dominated by the contribution from the collimated intensity, which can be expressed using the Beer-Lambert's law as $q_{detect} \doteq q_0 e^{-\beta L}$. It is noted that this equation cannot be directly used to determine the extinction coefficient since the effect of the cuvette wall is not considered in the above treatment.

A detailed analysis based on Monte Carlo simulation is also conducted to fully consider the multiple scattering in the medium and multiple reflections and refractions at liquid/glass/air interfaces under realistic experimental configuration. The total light radiation intensity is composed of a collimated and a diffuse component [27]. The total transmittance and the transmittance solely due to direct transmitted intensity were simulated by Monte Carlo method separately. Firstly, a fully simulation is used to obtain the total transmittance $T_{tot}$. Secondly, the transmittance due to the direct transmitted intensity ($T_{dir}$) is obtained by assuming the medium is non-scattering. The governing equations for the direct transmitted intensity and the diffuse intensity refer to Ref. [27]. The transmittance due to scattered photons ($T_{scat}$) can then be determined as $T_{scat} = T_{tot} - T_{dir}$. In the Monte Carlo simulation, the incident light beam had a diameter of 2 mm, which was incident perpendicularly on the cuvette (glass-sample-glass). A detector having diameter the same as the incident beam was used to record the transmitted photons. In the Monte Carlo analysis, the Henyey-Greenstein scattering phase function is used, the considered extinction coefficient is 100 m$^{-1}$, the thickness of glass and liquid layer is 1.5 and 3 mm, respectively, the asymmetric factor $g$ is 0.97, and the value of scattering albedo $\omega$ is 0.9. The half acceptance angle of detector is varied by changing the distance between the cuvette and detector.

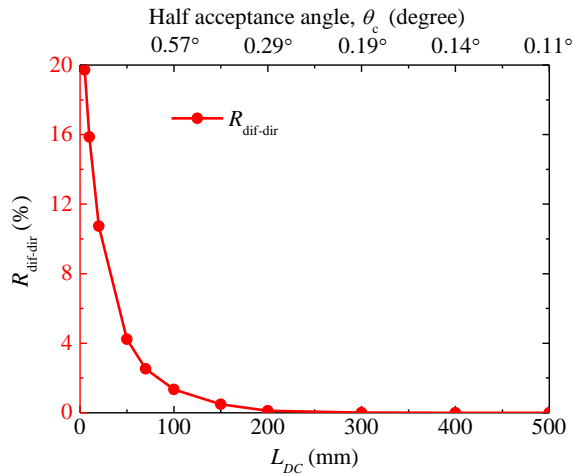

**Fig. 4.** The ratio of detected diffuse photons to the detected direct transmitted photons ($R_{dif-dir}$) at different detector-cuvette distance. The considered extinction coefficient is 100 m$^{-1}$, the thickness of glass and liquid layer is 1.5 and 3 mm, respectively, the asymmetric factor $g$ is 0.97, and scattering albedo $\omega$ is 0.9. The diameters of detector and incident light beam are 2 mm.

Figure 4 shows the simulated ratio of detected diffuse photons $q_{dif}$ to the detected direct transmitted photons $q_{dir}$, namely, $R_{dif-dir}$, at different detector-cuvette distance, which is evaluated as $R_{dif-dir} = |T_{scat} - T_{dir}|/T_{dir}$, in which $T_{scat}$ denotes the simulated transmittance considering multiple scattering and $T_{dir}$ denotes simulated transmittance only taking into account the direct transmitted photons. As shown, the $R_{dif-dir}$ decreases with decreasing half acceptance angle (increasing distance between the detector and the scattering medium). Since the scattering phase function of the particles is difficult to be accurately measured, especially for a wide range of spectra, it is expected that the term related to forward-scattering-peak correction can be neglected with specific experiment configuration during the measurement of extinction coefficient. This is true when the distance between the detector and the scattering medium is large enough, and then the half acceptance angle is extremely small. It is demonstrated that for the present experiment configuration, namely, the detector diameter is 2 mm and the distance between detector and cuvette is 200 mm, and the half acceptance angle $\theta_c$ = 0.29°. For the present experimental setup, $R_{dif-dir}$ is about 0.1%, indicating only a very small amount of diffuse photons are detected by the detector. As such, the forward-scattering-peak correction can be safely neglected for the present experimental setup. This analysis gives a reference of choosing proper detector-cuvette distance to simplify the measurement of extinction coefficient. It is noted that larger asymmetric factor will induce larger measurement uncertainty. The value of $g$ = 0.97 tested here will ensure that the above analysis covers particles with size of several tens of microns.

## 3. IMPROVED TRANSMISSION METHOD

Here, an improved transmission method is presented to determine the extinction coefficient of particle suspensions. The established model of light transmission and variables definition is shown in Fig. 1. In the following, relation of extinction coefficient with normal-normal transmittance with considering the higher-order transmission terms in the three-layer system is derived. The total transmittance of the system $T_{EXP}$ considering all the higher-order transmission terms can be expressed as [28, 29]

$$T_{EXP} = \frac{T_1 T_3 t_2}{1 - R_1' R_3 t_2^2} \qquad (11)$$

where $t_2 = e^{-\beta L_2}$, $T_1$ and $T_3$ represent the transmittance of layer 1 and layer 3, respectively, $R_1'$ and $R_3$ represent the reflectance of layer 1 from the non-incident side and layer 3 from the incident side, respectively. Equation (11) can be rewritten as

$$t_2^2 R_1' R_3 T_{EXP} + T_1 T_3 t_2 - T_{EXP} = 0 \qquad (12)$$

This is a quadratic equation of $t_2$. Hence the extinction coefficient of liquid-particle suspensions can be obtained as

$$\beta = -\frac{1}{L_2} \ln\left( \frac{-T_1 T_3 + \sqrt{T_1^2 T_3^2 + 4 T_{EXP}^2 R_1' R_3}}{2 T_{EXP} R_1' R_3} \right) \qquad (13)$$

The extinction coefficient of the particles can be calculated when the extinction coefficient of liquid-particle suspensions is obtained, $\beta_{particle} = \beta - \alpha_{base}$, where $\alpha_{base}$ is the absorption coefficient of the base fluid calculated from $\alpha_{base} = 4\pi\kappa_2/\lambda$. Note that the contribution from diffused light is omitted here following the study presented in Section 2. The transmittance of layer 1 and layer 3, and the reflectance of layer 1 from the non-incident side and layer 3 from the incident side are given as follows [30, 31]:

$$T_1 = \frac{\tau_{01}\tau_{12}e^{-\alpha_1 L_1}}{1 - \rho_{10}\rho_{12}e^{-2\alpha_1 L_1}}, \quad T_3 = \frac{\tau_{23}\tau_{30}e^{-\alpha_3 L_3}}{1 - \rho_{32}\rho_{30}e^{-2\alpha_3 L_3}} \qquad (14)$$

$$R_1' = \rho_{21} + \frac{\tau_{21}\tau_{12}\rho_{10}e^{-2\alpha_1 L_1}}{1 - \rho_{12}\rho_{10}e^{-2\alpha_1 L_1}}, \quad R_3 = \rho_{23} + \frac{\tau_{23}\tau_{32}\rho_{30}e^{-2\alpha_3 L_3}}{1 - \rho_{32}\rho_{30}e^{-2\alpha_3 L_3}} \qquad (15)$$

where $\rho_{ij}$, $\tau_{ij}$ represent the reflectivity and transmissivity at the interface between two neighboring media $i$ and $j$, which is defined as [30, 31]

$$\rho_{ij} = \frac{(n_j - n_i)^2 + (\kappa_j - \kappa_i)^2}{(n_j + n_i)^2 + (\kappa_j + \kappa_i)^2}, \quad \tau_{ij} = 1 - \rho_{ij} \qquad (16)$$

where $n_i + i\kappa_i$ represent the complex refractive index of the medium $i$; note that the subscript 0, 1, 2, and 3 denote the layer of air, left side glass, the base fluid and the right side glass, respectively. By using the measurement equation (Eq. (13)), the optical constants of the base fluid and glasses should be measured at first. Here, the double optical pathlength transmission method (DOPTM) [28, 29] is applied, which is suitable to measure the optical constants of liquids at a wide spectral range.

## 4. EXPERIMENTAL VALIDATION

We use the silicon dioxide particles with known optical constants and particle diameter distribution to verity the proposed improved transmission method. Based on Lorenz-Mie theory, the experimental validation of extinction coefficient of silicon dioxide microspheres was conducted for suspensions of various mean particle sizes.

Baseline Chromtech Research Centre, Tianjin, China) were chosen as exemplification particles, which was then dispersed into deionized water. The normal-normal transmittance were measured for liquid-particle suspensions and base fluid in cuvettes with thickness $L_2$ = 5 mm using the V–VASE ellipsometer (J.A. Woollam Company, USA) in the spectral range of 300–800 nm in the present configuration.

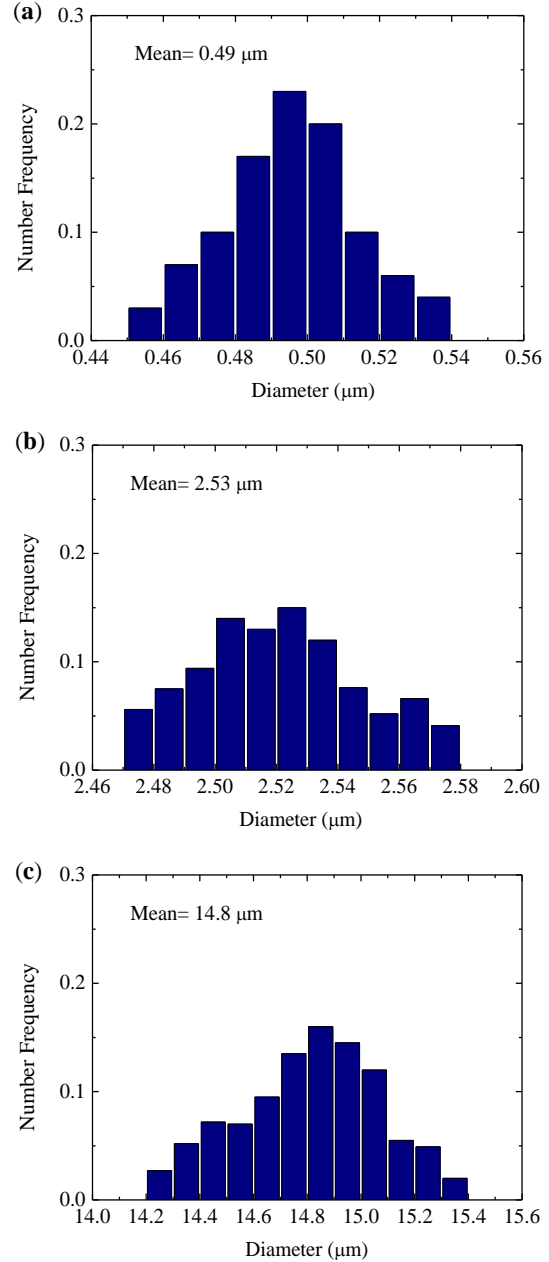

**Fig. 6.** Measured diameter distributions for silicon dioxide microspheres with three different mean diameters. (a) 0.49 μm, (b) 2.53 μm, and (c) 14.8 μm, respectively. Caption text with descriptions of (a), (b), and (c).

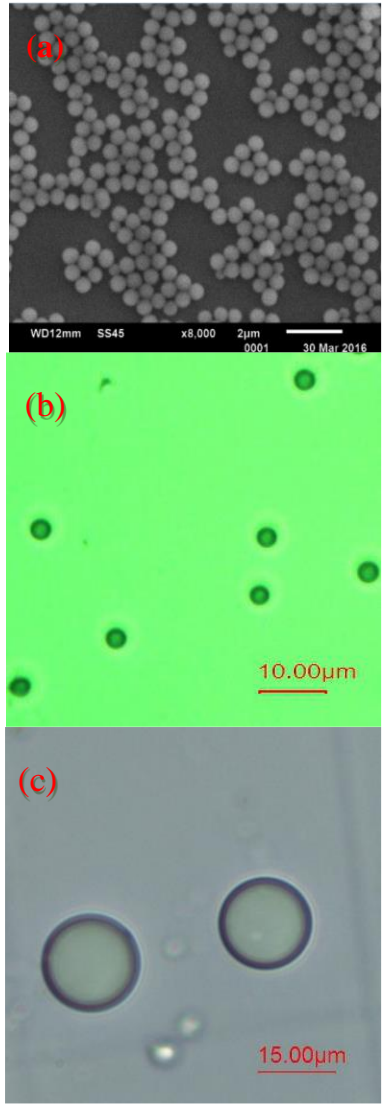

**Fig. 5.** Micrograph of silicon dioxide microspheres of three different mean diameters. (a) 0.49 μm, (b) 2.53 μm, and (c) 14.8 μm, respectively.

### A. Experimental setup and sample characteristics

The normal–normal transmittance of the sample was measured by using the V-VASE ellipsometer (with a wavelength resolution of 10 nm). The Xenon lamp (0.19–2 μm) was used as light source to cover the research wavelength. The V-VASE ellipsometer uses silicon and InGaAs photodiode detectors. The silicon dioxide microspheres (supplier:

The sample liquid suspensions was held in a cuvette with wall thickness of 1.2 mm. The distance between detector and cuvette is 200 mm. The base fluid was prepared by adding 0.2 wt% of dispersant (Chitosan) to distilled water. Note that the reliability of measured data is corresponding to the accuracy of transmittance measurement. According to our experience, more reliable results can be obtained if the measured transmittance is within the range of 0.1-0.85, which can be satisfied by adjusting the thickness of cuvette. Since the concentration of dispersant is very small, the optical property of base fluid has no observable difference from the deionized water. The diameters of detector and incident light beam are 2 mm. The optical constants of the glass of cuvette ($n_1 + i\kappa_1 = n_3 + i\kappa_3$) in 300–800 nm were measured by the V–VASE ellipsometer which have been reported in Refs. [28, 29]. The measured optical constants of deionized water were published in our earlier work [28]. All the measurements were carried out at room temperature and normal atmospheric pressure.

Figure 5 shows the micrograph of silicon dioxide microspheres with three different mean diameters, (a) 0.49 μm, (b) 2.53 μm, and (c) 14.8 μm, respectively. The micrograph (a) was obtained by scanning electron microscopy (JSM-6510MV, JEOL Ltd, Japan). The micrographs (b) and (c) were obtained using a biological microscope (UB203i-5.0M, China) connected to a CCD camera. Figure 6 shows the number frequency of the diameter of the silicon dioxide microspheres. In order to get a well-mixed sample, an ultrasonic oscillator (Shanghai Wuxiang, DL-1200D) was used to improve the dispersion of particles before experimental measurement. The particle size distributions were measured using a popular public domain image-viewing and processing program, namely, ImageJ software (developed at the National Institutes of Health, http://rsb.info.nih.gov/ij) after the transmittance measurement was finished. ImageJ reports the diameter distributions of the three samples of silicon dioxide microspheres with different mean diameters, (a) 0.49 μm, (b) 2.53 μm, and (c) 14.8 μm, respectively. More than 300 particles were counted for each sample. The particle concentration in each solution was counted using a Petroff-Hausser counting chamber (Hausser scientific, USA).

**B. Experimental uncertainty**

In the improved transmission method, the experimental uncertainty of extinction coefficient is obtained by Eq. (13). The transmittance of each sample was measured six times. The average spectral transmittance is denoted as $\overline{T}_{EXP}$ and the related standard deviation of arithmetic mean $\sigma_{\overline{T}_{EXP}}$ can be expressed as

$$\overline{T}_{EXP} = \frac{1}{M}\sum_1^M T_M \quad (17)$$

$$\sigma_{\overline{T}_{EXP}} = \sqrt{\frac{1}{M \times (M-1)}\sum_1^M (T_M - \overline{T}_{EXP})^2} \quad (18)$$

where $M$ = 6 denotes the number of measurements. Then, the combined standard uncertainty of the measured transmittance $\Delta T_{EXP}$ at a confidence level of 68% is calculated from

$$\Delta T_{EXP} = \sqrt{(1.11\sigma_{\overline{T}_{EXP}})^2 + \Delta^2} \quad (19)$$

where $\Delta$ denotes the uncertainty of the transmission measurement, which is directly given by the instrument at each wavelength.

Note that the extinction coefficient, which is determined based on these related parameters, can be written in function form as

$$\beta = f(n_1, n_2, n_3; \kappa_1, \kappa_2, \kappa_3; L_1, L_2, L_3; T_{EXP}) \quad (20)$$

The uncertainty of the extinction coefficient can be computed as

$$(\Delta \beta)^2 = \sum_i \left[\frac{\partial \beta}{\partial \chi_i} \cdot \Delta \chi_i\right]^2 \quad (21)$$

where $\Delta \chi_i$ represents the uncertainty of the different related parameters in Eq. (20). The uncertainties of thickness ($\Delta L_1$, $\Delta L_2$ and $\Delta L_3$) measurements are 0.1 mm. The uncertainty of the particle extinction coefficient $\Delta \beta_{particle}$ is calculated from

$$|\Delta \beta_{particle}| = |\Delta \beta| + |\Delta \beta_{base}| \quad (22)$$

Based on Eq. (3), for the traditional transmission method, the experimental uncertainty of extinction coefficient is obtained from

$$(\Delta \beta_{particle}^{trad})^2 = \left[\frac{\partial \beta_{particle}^{trad}}{\partial L_2}\Delta L_2\right]^2 + \left[\frac{\partial \beta_{particle}^{trad}}{\partial T_{EXP}}\Delta T_{EXP}\right]^2 + \left[\frac{\partial \beta_{particle}^{trad}}{\partial T_{EXP,Ref}}\Delta T_{EXP,Ref}\right]^2 \quad (23)$$

The uncertainty of the transmittance measurement is about 1.4%, which is directly given by V-VASE ellipsometer.

**C. Experimental results**

Previous studies suggested that independent scattering should occur at small particle volume fraction, viz, $f_v < 0.006$ (interparticle clearance measured in wavelength c/λ>0.5) [27, 32]. In this paper, the particle volume fractions of all the sample particle suspensions considered are very small, $f_v < 0.0004$ (c/λ>209). Therefore, light scattering by particles is independent and the Lorenz-Mie theory is employed in the following to predict the radiative properties of liquid-particle suspensions.

Figure 7 shows the measured extinction coefficient of silicon dioxide microspheres by the improved and the traditional transmission methods. The measured data are compared with predicted values based on Lorenz-Mie theory. In the Lorenz–Mie theory analysis, the complex refractive index of silicon dioxide microspheres is taken from Ref. [33], and the particle size distributions are obtained experimentally as shown in Fig. 6. The asymmetric factors $g$ of the three sample of silicon dioxide microspheres (with mean diameters of 0.49 μm, 2.53 μm, and 14.8 μm) are about 0.90~0.96, 0.89~0.93, and 0.95~0.96, respectively, in the research spectral range. As shown in Fig. 7, the measured results obtained by the improved method agree well with the predicted values of Lorentz-Mie theory. Differences in extinction coefficient between the improved and the traditional transmission methods can be observed. The smaller the optical thickness $\beta_{particle}L_2$ of particles, the bigger the relative difference of the extinction coefficients obtained by these two methods is. This result agrees with the numerical analysis presented in Section 1.

For the sake of analysis, the relative error of measured extinction coefficient of particles is defined as

$$\Delta E = \left|\frac{\beta_{EXP} - \beta_{Mie}}{\beta_{Mie}}\right| \times 100\% \quad (24)$$

where $\beta_{EXP}$ stands for the measured extinction coefficient, $\beta_{Mie}$ represents the predicted values based on Lorenz-Mie theory. Figure 8 shows the relative errors of the measured extinction coefficients of the improved and the traditional transmission methods. The relative errors of extinction coefficient for the traditional transmission method are around 2.5~59%, 1.1~9% and 2.7~13% with different conditions of (a) $D$ = 0.49 μm, $N$ = 4.62×10$^{14}$ particles/m$^3$, (b) $D$ = 2.53 μm, $N$ = 3.89×10$^{13}$ particles/m$^3$ and (c) $D$ = 14.8 μm, $N$ = 2.38×10$^{11}$ particles/m$^3$, respectively. The relative errors of extinction coefficient for the improved transmission method is much less than that of the traditional transmission method, which are about 0.1~10%, 0.4~4% and 0.3~3% respectively for these three conditions stated above. Generally, the improved transmission method is more accurate than the traditional transmission method.

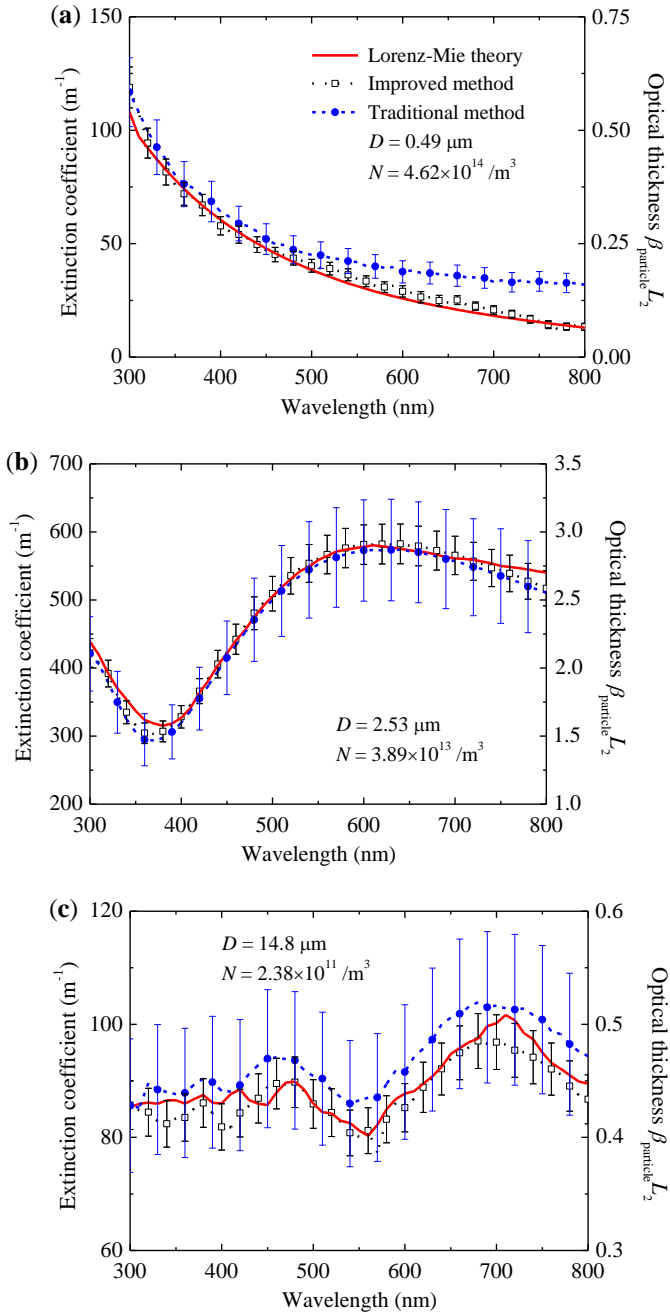

**Fig. 7.** The extinction coefficient of silicon dioxide microspheres obtained by Lorenz–Mie theory predictions and the data determined by the improved and the traditional transmission methods. The mean diameters $D$ of silicon dioxide microspheres are equal to about (a) 0.49 μm, (b) 2.53 μm, and (c) 14.8 μm, respectively. The concentration $N$ of silicon dioxide microspheres are (a) $4.62 \times 10^{14}$ particles/m³, (b) $3.89 \times 10^{13}$ particles/m³, and (c) $2.38 \times 10^{11}$ particles/m³, respectively.

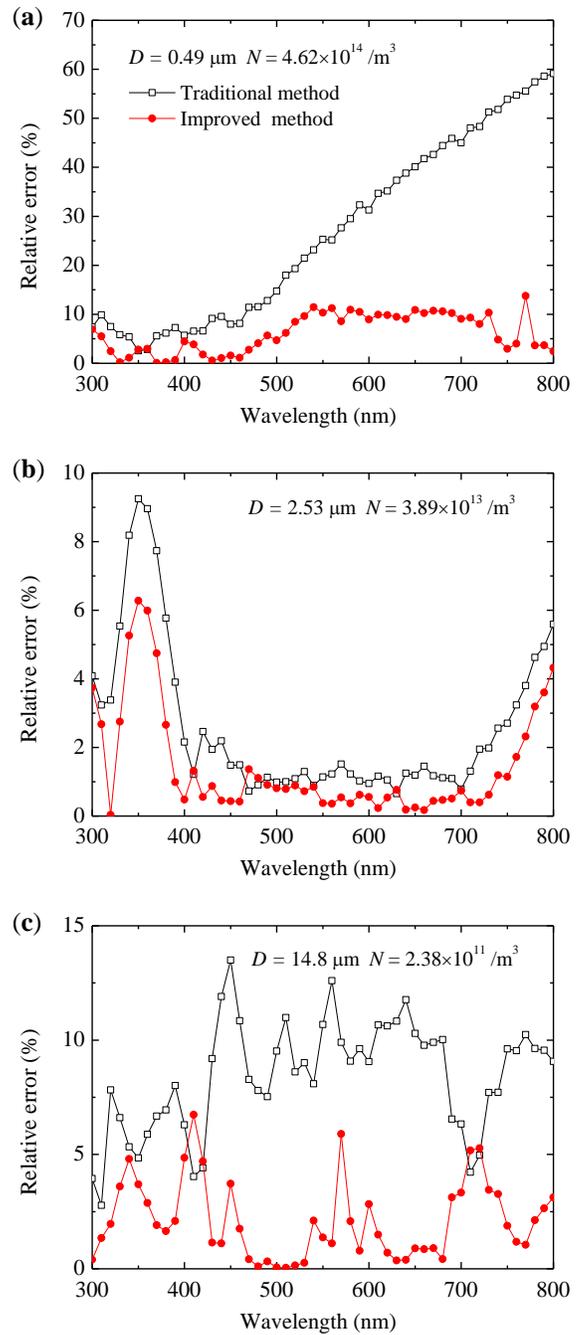

**Fig. 8.** The relative errors from true value using the improved transmission method and the traditional transmission method.

## 5. CONCLUTION

An improved transmission method is proposed to measure the extinction characteristics of micro/nano particle suspensions. The effects of multiple reflections and refractions at glass-liquid/glass-air interfaces are taken into account in the new method. Both theoretical and experimental analysis prove that the improved method we propose is more accurate than the traditional one. The refractive index of glass is an important factor that influence the measurement accuracy of extinction coefficient of particles. In the traditional transmission method, the error caused by omitting higher-order transmission terms becomes large and cannot be neglected when the difference of the refractive index between the glass and sample (or glass and air) is increased. The forward-scattering-peak correction is demonstrated to be negligible with small detector acceptance angle (e.g., half acceptance angle less than 0.3° even for asymmetry factor g=0.97). This will greatly simplify the measurement of extinction coefficient since the scattering phase function needs not to be determined in advance.

The extinction coefficient of silicon dioxide microspheres with known optical constants and particle diameter distribution were measured by the improved and the traditional transmission methods in spectral range from 300 nm to 800 nm and were compared with predicted values based on Lorenz-Mie theory. For the traditional transmission method, relative error is bigger than the improved method for the case with small optical thickness of particle suspensions. The results show that the higher-order transmission terms caused by multiple reflections at medium interfaces cannot be ignored when the optical thickness of particles is small. Generally, the improved transmission method is demonstrated to have good accuracy in the measurement of extinction coefficient of liquid suspensions containing nano, submicron, and/or micron particles, even if the internal scattering is prominent and strongly forward scattering. This provides an alternative and more accurate way for measuring the extinction characteristics of micro/nano particle suspensions.

**Funding.** The supports by National Nature Science Foundation of China (Nos. 51336002, 51276051, 51421063) are greatly acknowledged. JMZ also thanks the support by the Fundamental Research Funds for the Central Universities (Grant No. HIT. NSRIF. 2013094, HIT.BRETIII.201415).